\documentclass[prl,aps,floatfix,twocolumn,superscriptaddress,preprintnumbers]{revtex4-1}

\usepackage{amsmath,amsfonts,amssymb,mathtools,cases,slashed,bm}
\usepackage{mathrsfs}
\usepackage{color}
\usepackage[dvipsnames]{xcolor}
\usepackage[colorlinks=true,citecolor=blue,linkcolor=blue,urlcolor=blue]{hyperref}
\usepackage{tabularx}
\usepackage{physics}
\usepackage[normalem]{ulem}
\usepackage{epsfig, graphicx}
\usepackage{hhline}

\usepackage{lineno}
\newcommand{\fref}[1]{Fig.~\ref{#1}}

\newcommand{\eref}[1]{Eq.~(\ref{#1})}

\newcommand{\tref}[1]{Table~\ref{#1}}

\newcommand{\diff}[1]{\mathrm{d}#1}

\newcommand{\hel}{{^3 {\rm He}}}

\newcommand{\chidof}{{\chi^2/N_{\rm dat}}}
\newcommand{\pub}{{\Delta \bar{u}}}
\newcommand{\pdb}{{\Delta \bar{d}}}

\begin{document}

\title{Polarized Antimatter in the Proton from Global QCD Analysis}

\author{C. Cocuzza}
\affiliation{Department of Physics, SERC, Temple University, Philadelphia, Pensylvannia 19122, USA}
\author{W. Melnitchouk}
\affiliation{Jefferson Lab, Newport News, Virginia 23606, USA \\
        \vspace*{0.2cm}
        {\bf Jefferson Lab Angular Momentum (JAM) Collaboration
        \vspace*{0.2cm} }}
\author{A. Metz}
\affiliation{Department of Physics, SERC, Temple University, Philadelphia, Pensylvannia 19122, USA}
\author{N. Sato}
\affiliation{Jefferson Lab, Newport News, Virginia 23606, USA \\
        \vspace*{0.2cm}
        {\bf Jefferson Lab Angular Momentum (JAM) Collaboration
        \vspace*{0.2cm} }}

\begin{abstract}
We present the first simultaneous global QCD analysis of spin-dependent parton distribution functions alongside their spin-averaged counterparts and pion, kaon, and unidentified hadron fragmentation functions.  This analysis includes all data relevant for constraining the polarized light quark sea asymmetry $\Delta \bar{u} - \Delta \bar{d}$, in particular the latest polarized $W$-lepton production data from the STAR collaboration at RHIC and semi-inclusive deep inelastic scattering data from COMPASS, allowing the most robust extraction available with minimal theoretical assumptions.  We also extract a self-consistent set of antiquark polarization ratios $\Delta \bar{u}/\bar{u}$ and $\Delta \bar{d}/\bar{d}$ and determine the signs of the truncated contributions to the proton spin from the light 
antiquarks.
\end{abstract}

\date{\today}
\preprint{JLAB-THY-22-3562}
\maketitle

{\it Introduction.---}\ 
Understanding the detailed decomposition of the proton spin into its constituent quark and gluon helicity and orbital angular momentum components promises to be one of the most significant accomplishments in nuclear and particle physics of this generation~\cite{Aidala:2012mv, Jimenez-Delgado:2013sma, Leader:2013jra}.
While the total light quark contributions to the helicity are well determined from polarized inclusive deep-inelastic scattering (DIS) data~\cite{EuropeanMuon:1989yki, SpinMuon:1998eqa, SpinMuon:1999udj, COMPASS:2006mhr, COMPASS:2010wkz, COMPASS:2015rvb, Baum:1983ha, E142:1996thl, E154:1997xfa, E143:1998hbs, E155:1999pwm, E155:2000qdr, HERMES:1997hjr, HERMES:2006jyl}, and jet production in polarized $pp$ collisions \cite{STAR:2006opb, STAR:2007rjc, STAR:2012hth, STAR:2014wox, STAR:2019yqm, STAR:2021mqa, STAR:2021mfd, PHENIX:2010aru} provides constraints on the gluon helicity \cite{deFlorian:2014yva, Zhou:2022wzm}, far less is known about the polarization of the antiquark sea.
There have been some intriguing hints of a polarized antiquark asymmetry, $\pub-\pdb$, from polarized semi-inclusive DIS (SIDIS) measurements \cite{SpinMuon:1997yns, HERMES:2004zsh, HERMES:1999uyx, COMPASS:2010hwr, COMPASS:2009kiy}, in analogy with the spin-averaged $\bar{u}-\bar{d}$ asymmetry inferred from unpolarized DIS and Drell-Yan measurements~\cite{NewMuon:1991hlj, NewMuon:1993oys, NuSea:2001idv, NuSea:1998kqi, SeaQuest:2021zxb}.
Various nonperturbative model calculations have also been performed~\cite{Schreiber:1991tc, Cao:2003zm, Diakonov:1997, Wakamatsu:1999vf, Bourrely:2015kla}, some of which predict~\cite{Diakonov:1997, Wakamatsu:1999vf, Bourrely:2015kla} large positive $\pub-\pdb$ asymmetries.

Recently more probes of antiquark polarization have been possible through  $W$-lepton production in polarized $pp$ collisions.
In particular, the STAR \cite{STAR:2010xwx, STAR:2014afm, STAR:2018fty} and PHENIX \cite{PHENIX:2015ade, PHENIX:2018wuz} collaborations at RHIC have used polarized $pp$ collisions at center of mass energy $\sqrt{s} = 510$~GeV to measure the longitudinal single--spin asymmetry 
    \mbox{$A_L = (\sigma_+ - \sigma_-)/(\sigma_+ + \sigma_-)$}, 
where $\sigma_+$ ($\sigma_-$) is the cross section for positive (negative) proton helicity, for the leptonic decay channels $W^+ \to e^+ \nu$ and $W^- \to e^-\bar{\nu}$.
At leading order, these can be written as 
\begin{subequations}
\begin{eqnarray}
A_L^{W^+} \propto 
\frac{\Delta \bar{d}(x_1) u(x_2) - \Delta u(x_1) \bar{d}(x_2)}
     {\bar{d}(x_1) u(x_2) + u(x_1) \bar{d}(x_2)},
\\
\notag
\\
A_L^{W^-} \propto 
\frac{\Delta \bar{u} (x_1) d(x_2) - \Delta d(x_1) \bar{u}(x_2)}
     {\bar{u}(x_1) d(x_2) + d(x_1) \bar{u}(x_2)},
\end{eqnarray}
\label{e.ALW}%
\end{subequations}%
where $\Delta f$ ($f$) represents a polarized (unpolarized) parton distribution function (PDF) evaluated at momentum fraction $x_1$ ($x_2$) carried by the parton in the polarized (unpolarized) proton.
Combined with 
DIS observables, these asymmetries provide a vital new handle on the extraction of the polarized antiquark PDFs $\pub$ and $\pdb$.

Previous global analyses~\cite{deFlorian:2009vb, deFlorian:2014yva, DeFlorian:2019xxt, Nocera:2014gqa} have sought to extract the $\pub-\pdb$ asymmetry under various assumptions and with different methods for estimating uncertainties.
De~Florian {\it et al.} (DSSV) \cite{deFlorian:2009vb} extracted a positive $\pub-\pdb$ from spin-dependent data with fixed input for unpolarized PDFs and fragmentation functions (FFs), assuming PDF positivity and SU(3) symmetry for axial-vector charges within errors.
The Monte Carlo analysis by the NNPDF collaboration~\cite{Nocera:2014gqa} generated prior samples from the DSSV fit~\cite{deFlorian:2009vb}, 
thus inheriting the corresponding assumptions.
The NNPDF analysis also used a reweighting procedure involving $\chi^2$-based weights, which is inconsistent with the Gaussian likelihood used in the generation of the replicas~\cite{Sato:2013ika}.

Instead of relying on reweighting prescriptions and assumptions about PDF positivity or flavor symmetry, here we present a new simultaneous global QCD analysis of unpolarized and polarized PDFs and FFs, including for the first time STAR $A_L^{W}$ data, along with data on inclusive and semi-inclusive polarized lepton-nucleon DIS, unpolarized SIDIS, and jet production in polarized $pp$ collisions~\cite{Zhou:2022wzm}.
The simultaneous Monte Carlo analysis allows us to more reliably quantify the uncertainties on all distributions, and examine the interplay between the sea asymmetry and parametrizations of FFs.
The simultaneous determination of both types of PDFs also provides the first self-consistent extraction of the antiquark polarization ratios $\Delta \bar{u}/\bar{u}$ and $\Delta \bar{d}/\bar{d}$.

{\it Theoretical framework.---}\ 
Our theoretical framework is based on fixed-order collinear factorization for high-energy scattering processes, including DIS, Drell-Yan lepton-pair production, and weak boson and jet production in hadronic collisions.
The single-spin asymmetry $A_L^{W}$ has unique sensitivity to both unpolarized and polarized PDFs, giving further motivation for performing a simultaneous analysis of both types of PDFs.
The cross section for this process can be written as differential in the lepton pseudorapidity, $\eta_\ell$, and its transverse momentum, $p_T^\ell$.
The renormalization and factorization scales are chosen to be the mass of the $W$ boson, $\mu_R = \mu_F = M_W$, and the NLO expressions for the hard scattering kernels are found in Ref.~\cite{Ringer:2015oaa}.

The scale dependence of the PDFs is determined according to the DGLAP evolution equations~\cite{Gribov:1972ri, Dokshitzer:1977sg, Altarelli:1977zs}, with the PDFs and $\alpha_s$ evolved at next-to-leading logarithmic accuracy with the boundary condition $\alpha_s(M_Z)=0.118$. 
For light as well as heavy quarks the PDFs are evolved using the zero-mass variable flavor number scheme.
The values of the heavy quark mass thresholds for the evolution are taken from the PDG as $m_c=1.28$~GeV and $m_b=4.18$~GeV in the $\overline{\textrm{MS}}$ scheme~\cite{ParticleDataGroup:2018ovx}.

Our PDF extraction procedure is based on Bayesian inference using the Monte Carlo techniques developed in previous JAM analyses~\cite{Sato:2016tuz, Sato:2016wqj, Ethier:2017zbq, Sato:2019yez, Moffat:2021dji}.
The parameterization of the unpolarized PDFs is discussed in Ref.~\cite{Cocuzza:2021cbi}, while for the polarized PDFs and FFs at the input scale $\mu_0 = m_c$ we use the form
\begin{eqnarray}
f(x,\mu_0) = N x^\alpha (1-x)^\beta (1 + \eta x),
\label{e.template}
\end{eqnarray}
where $N$, $\alpha$, $\beta$, and $\eta$ are fit parameters.
The polarized light quark PDFs $\Delta u$ and $\Delta d$ are parameterized as a sum of a valence and a sea component.
For the sea quark $\Delta \bar{u}$, $\Delta \bar{d}$, $\Delta s$, and $\Delta \bar{s}$ PDFs we use two functions of the form~(\ref{e.template}), one of which is unique to each flavor while the other describes the low-$x$ region and is shared between all four distributions.

The same template (\ref{e.template}) is used for FFs, but with $x$ replaced by the momentum fraction $z$ of the parton carried by the hadron, and with $\eta = 0$.
For the $\pi^+$ FFs, we assume charge symmetry,
    $D^{\pi^+}_u = D^{\pi^+}_{\bar{d}}$,
    $D^{\pi^+}_d = D^{\pi^+}_{\bar{u}}$, 
as well as 
    $D^{\pi^+}_q = D^{\pi^+}_{\bar{q}}$ for heavier quarks $q = s, c, b$,
while for the $K^+$ FFs we take
    $D^{K^+}_d = D^{K^+}_{\bar{u}} = D^{K^+}_{\bar{d}}$
and
    $D^{K^+}_q = D^{K^+}_{\bar{q}}$ for $q = c, b$,
but allow the favored $D^{K^+}_u$ and $D^{K^+}_{\bar{s}}$ FFs to differ.
The FFs for negatively charged mesons are related by
    $D_q^{\pi^-/K^-} = D_{\bar{q}}^{\pi^+/K^+}$ for all flavors.
We use two shapes each for $D_u^{\pi^+}$, $D_d^{\pi^+}$, $D_u^{K^+}$, and $D_d^{K^+}$, and one shape for all other quark and gluon FFs.
The parametrizations for the unidentified hadron FFs are identical to those in Ref.~\cite{Moffat:2021dji}.
We tested that adding further flexibility to the FFs, such as $\eta \neq 0$, does not affect the quality of the fit nor the extracted distributions.
Overall, 35 leading-twist PDFs and FFs are fitted with a total of 146 parameters.
Including parameters for higher-twist and off-shell corrections to structure functions, plus data normalizations, brings the number of parameters to~227.

Recently the question of PDF positivity beyond leading order in $\alpha_s$ in the $\overline{\rm MS}$ scheme has been debated~\cite{Candido:2020yat, Collins:2021vke}.
Such a constraint would require $|\Delta f(x,Q^2)| \leq f(x,Q^2)$ to hold for all flavors at all $x$ and $Q^2$.
To explore this question phenomenologically, we perform analyses with and without the positivity constraints.
The baseline analysis, referred to in the following as ``JAM'', does not enforce positivity; however, when included, the positivity constraints are enforced approximately on each Monte Carlo replica by imposing a penalty on the $\chi^2$ function when the bounds are violated~\cite{Ball:2013lla}.

\begin{table}[b]
\caption{Summary of $\chi^2$ values per number of points $N_{\rm dat}$ for the various datasets used in this analysis.}
\begin{tabular}{l r c}
\hhline{===}
process                 & $N_{\rm dat}$      & ~~~~~~~~$\chidof$       \\ 
\hline
\textbf{polarized} & &  \\
~~~~inclusive DIS               & 365        & ~~~~~~~~0.95  \\
~~~~SIDIS ($\pi^+,\pi^-$)       &  64        & ~~~~~~~~1.05  \\
~~~~SIDIS ($K^+,K^-$)           &  57        & ~~~~~~~~0.42  \\
~~~~SIDIS ($h^+,h^-$)           & 110        & ~~~~~~~~0.95  \\
~~~~inclusive jets              &  83        & ~~~~~~~~0.84  \\ 
~~~~STAR $W^{\pm}$              &  12        & ~~~~~~~~0.65  \\
~~~~PHENIX $W^{\pm}/Z$          &   6        & ~~~~~~~~0.50  \\
~~~~\textbf{total}              & {\bf 697}  & ~~~~~~~~{\bf 0.89} \\
\hline
\textbf{unpolarized} & & \\
~~~~inclusive DIS               & 3908       & ~~~~~~~~1.17  \\
~~~~SIDIS ($\pi^+,\pi^-$)       &  498       & ~~~~~~~~0.94  \\
~~~~SIDIS ($K^+,K^-$)           &  494       & ~~~~~~~~1.31  \\
~~~~SIDIS ($h^+,h^-$)           &  498       & ~~~~~~~~0.71  \\
~~~~inclusive jets              &  198       & ~~~~~~~~1.28  \\
~~~~Drell-Yan                   &  205       & ~~~~~~~~1.21  \\
~~~~$W$/$Z$ production~~~~~~~   &  153       & ~~~~~~~~1.01  \\
~~~~\textbf{total}              & {\bf 5954} & ~~~~~~~~{\bf 1.12} \\
\hline
SIA ($\pi^{\pm}$)               &  231       & ~~~~~~~~0.91 \\
SIA ($K^{\pm}$)                 &  213       & ~~~~~~~~0.70 \\
SIA ($h^{\pm}$)                 &  120       & ~~~~~~~~1.07 \\
\hline
\textbf{total}                  & {\bf 7215} & ~~~~~~~~{\bf 1.08} \\
\hhline{===}
\end{tabular}
\label{t.chi2}
\end{table}

{\it Quality of fit.---}\
Our analysis includes measurements of the DIS asymmetries $A_{\parallel}$ and $A_1$ for the proton, deuteron, and $\hel$ from 
EMC \cite{EuropeanMuon:1989yki}, 
SMC \cite{SpinMuon:1998eqa, SpinMuon:1999udj}, 
COMPASS \cite{COMPASS:2006mhr, COMPASS:2010wkz, COMPASS:2015rvb}, 
SLAC \cite{Baum:1983ha, E142:1996thl, E154:1997xfa, E143:1998hbs, E155:1999pwm, E155:2000qdr}, 
and HERMES \cite{HERMES:1997hjr, HERMES:2006jyl}.
To ensure the asymmetries are dominated by the leading-twist $g_1$ structure function, with negligible contributions from $g_2$, we restrict the four-momentum transfer squared to $Q^2 > m_c^2$ and the hadronic final state masses to $W^2 > 10$~GeV$^2$.
With the same cuts we include pion, kaon, and unidentified hadron SIDIS measurements on polarized proton, deuteron, and $^3$He targets from 
HERMES \cite{HERMES:2004zsh,HERMES:1999uyx}, COMPASS \cite{COMPASS:2009kiy, COMPASS:2010hwr} and SMC \cite{SpinMuon:1997yns}, with the fragmentation variable restricted to $0.2 < z < 0.8$ to ensure the applicability of the leading-power formalism and avoid threshold corrections~\cite{Moffat:2021dji}.

\begin{figure}[t]
\includegraphics[width=0.48\textwidth]{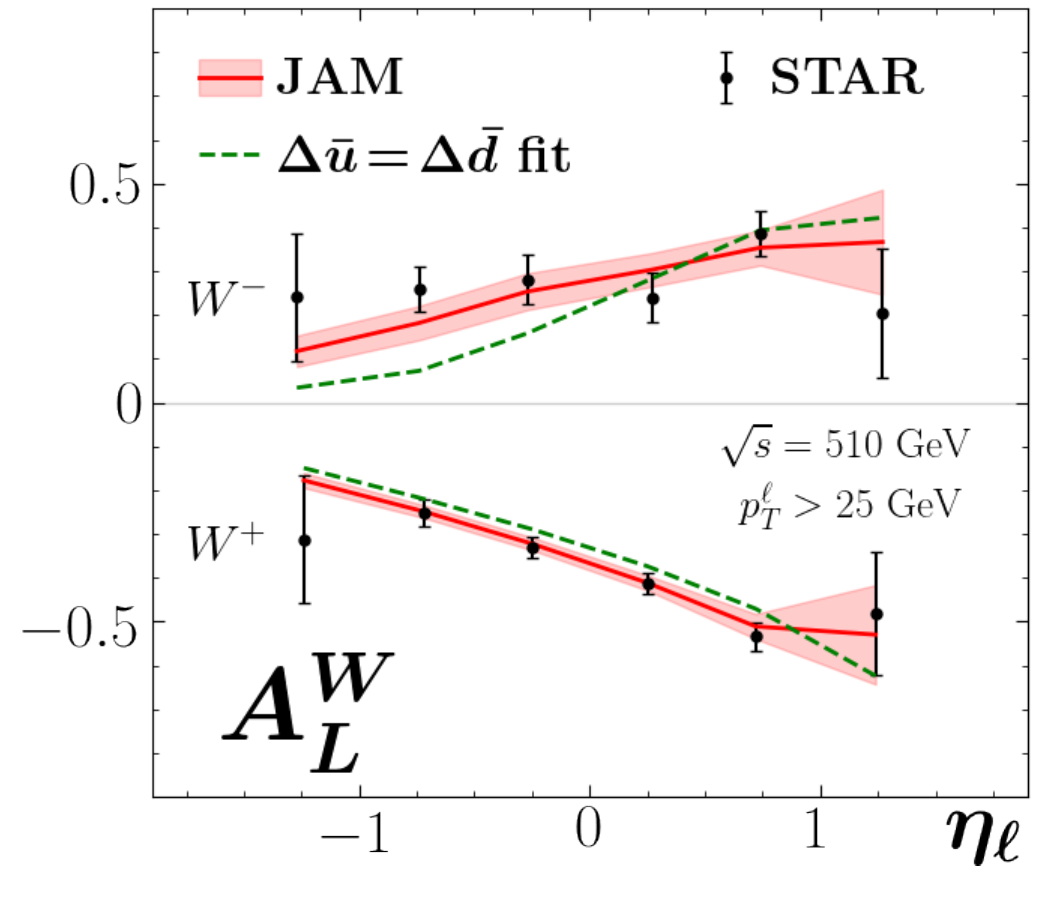}
\caption{Single-spin asymmetries $A_L^W$ versus pseudorapidity $\eta_\ell$ from STAR \cite{STAR:2018fty} (black circles) at $ \sqrt{s}=510$~GeV and integrated over $p_T^\ell > 25$~GeV, compared with the full JAM fit (red solid lines and $1\sigma$ uncertainty bands) and with a fit where $\pub$ is set equal to $\pdb$ (green dashed lines).}
\label{f.star}
\end{figure}

Beyond polarized lepton scattering, we describe jet production data in polarized $pp$ collisions from STAR~\cite{STAR:2006opb, STAR:2007rjc, STAR:2012hth, STAR:2014wox, STAR:2019yqm, STAR:2021mqa, STAR:2021mfd} and PHENIX~\cite{PHENIX:2010aru}, with a cut on the jet transverse momentum of 8~GeV~\cite{Zhou:2022wzm}.
We also include for the first time single-spin asymmetry $A_L^{W}$ data from STAR~\cite{STAR:2018fty} and $A_L^{W/Z}$ from PHENIX~\cite{PHENIX:2015ade, PHENIX:2018wuz}, which provide the most direct constraints on the antiquark polarization.
For unpolarized processes, we use data from inclusive DIS, Drell-Yan lepton-pair production, and inclusive $W^\pm$, $Z$ and jet production in hadronic collisions, as in Ref.~\cite{Cocuzza:2021rfn}.
To further constrain the FFs we also utilize SIA and unpolarized SIDIS data on pions, kaons, and unidentified hadrons, as in Ref.~\cite{Moffat:2021dji}.

The quality of our global analysis is summarized in \tref{t.chi2}, which shows a global $\chidof=1.08$ for $N_{\rm dat}=7215$ data points (697 for polarized, 5954 for unpolarized, and 564 for SIA).
The $\chidof$ for each experiment is generally stable whether PDF positivity constraints are imposed or not.
When enforcing $\pub = \pdb$, there are significant increases in
$\chidof$ for the STAR $W$ data (from 0.65 to 3.00), PHENIX $W/Z$ data at mid rapidity (0.21 to 2.23), \mbox{COMPASS} $A_{1p}^{\pi^-}$ data (1.19 to 1.60, as observed in Ref.~\cite{Ethier:2017zbq}), and SMC $A_{1p}^{h^-}$ data (1.10 to 1.53).
The STAR $A_L^{W}$ data are compared with the JAM fit in \fref{f.star} versus the pseudorapidity $\eta_\ell$.
When the asymmetry is forced to vanish, the quality of the fit suffers the most for $A_L^{W^-}$ at low~$\eta_\ell$.
This can be understood from \eref{e.ALW}, which shows that the asymmetries are most sensitive to $\pub$ and $\pdb$ at backward rapidity, where the first terms dominate due to $x_2$ being large and thus $q(x_2) \gg \bar{q}(x_2)$ for $q=u,d$.

\begin{figure}[t]
\includegraphics[width=0.48\textwidth]{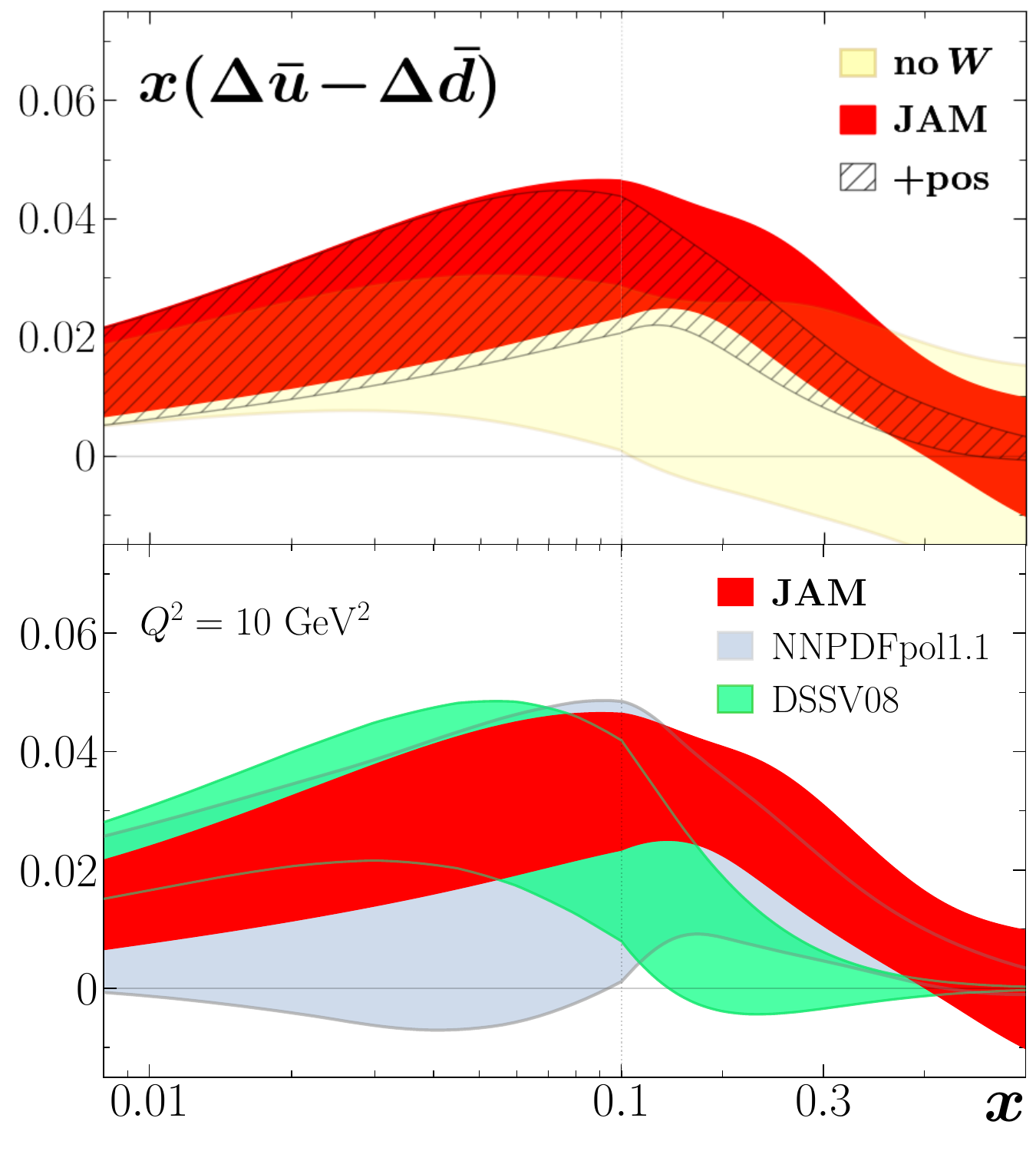}
\caption{Polarized sea quark asymmetry $x(\pub-\pdb)$ from JAM (red 1$\sigma$ bands) at $Q^2=10$~GeV$^2$ compared with: [top panel] fit without RHIC $W/Z$ data (yellow band) and the result with positivity constraints (hatched band), and [bottom panel] the NNPDFpol1.1 \cite{Nocera:2014gqa} and DSSV08 \cite{deFlorian:2009vb} analyses.}
\label{f.asym}
\end{figure}

{\it QCD analysis.---}\
The extracted unpolarized PDFs are nearly identical to those from the recent unpolarized JAM analyses~\cite{Cocuzza:2021cbi, Cocuzza:2021rfn}, while the FFs are consistent with those from Ref.~\cite{Moffat:2021dji}.
In this work we focus on the polarized PDFs, extracted from an
analysis of over 1,000 Monte Carlo samples.
The polarized antiquark asymmetry is shown in \fref{f.asym} and indicates a clear nonzero sea asymmetry for $0.01 < x < 0.3$.
The inclusion of positivity constraints significantly reduces the uncertainties at $x \gtrsim 0.2$, since the polarized sea quarks are restricted by the size of the unpolarized sea quarks.
In contrast to the final result, the results without the RHIC $W$ data are consistent with zero for $x \gtrsim 0.07$, illustrating the importance of the STAR $W$ data for the extraction of the polarized antiquark asymmetry in the intermediate-$x$ region.

In Fig.~\ref{f.asym} we also compare our results to the asymmetries from the DSSV~\cite{deFlorian:2009vb} and NNPDF~\cite{Nocera:2014gqa} groups.
The DSSV fit~\cite{deFlorian:2009vb} is qualitatively similar to our result without the RHIC $W$ data, as expected, with significantly smaller errors at high $x$ due to the inclusion of positivity constraints.
The differences in the shape 
at $x \lesssim 0.1$ may be attributable to the propagation of FF uncertainties and polarized PDF parametrization choice.

The NNPDF result~\cite{Nocera:2014gqa}, on the other hand, shows only a slight deviation from zero at high values of $x$.
This is consistent with this 
fit taking the DSSV result~\cite{deFlorian:2009vb} as the prior for $\pub$ and $\pdb$, but with 4$\sigma$ uncertainty, and including 
the older STAR $W$ data \cite{STAR:2014afm} in their reweighting analysis.
Our analysis is thus the first extraction of a nonzero polarized antiquark asymmetry in the intermediate-$x$ region, where model calculations generally indicate the largest effects~\cite{Diakonov:1997, Wakamatsu:1999vf, Bourrely:2015kla}.

\begin{figure}[t]
\includegraphics[width=0.48\textwidth]{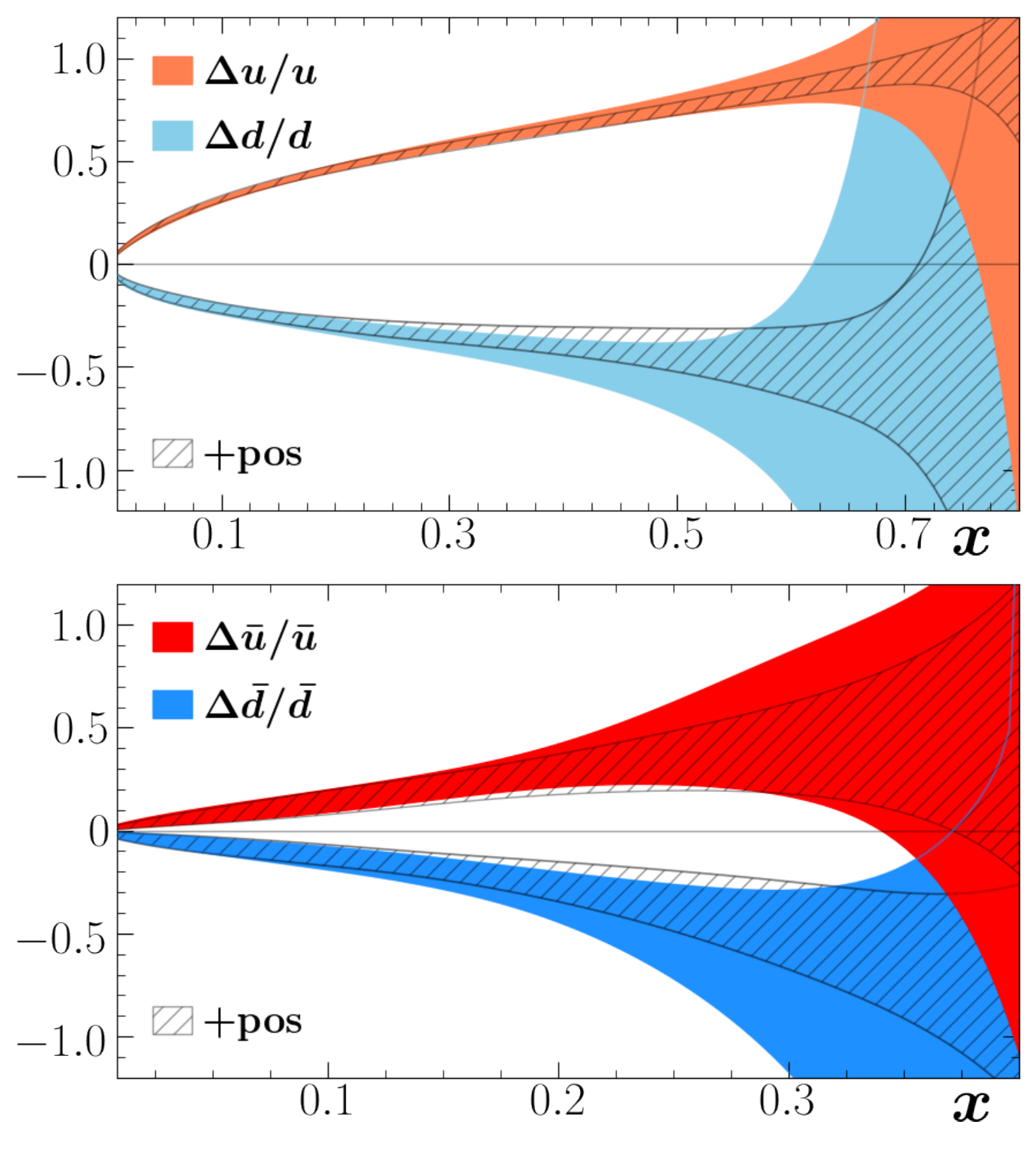}
\caption{Light sea quark polarization ratios $\Delta q/q$ at $Q^2 = 10$~GeV$^2$: [top panel] $u$ and $d$ (coral and skyblue 1$\sigma$ bands),
[bottom panel] $\bar{u}$ and $\bar{d}$ (red and blue 1$\sigma$ bands), compared with results with positivity constraints (hatched bands).}
\label{f.pol}
\end{figure}

The results for the light quark polarization ratios $\Delta q/q$ are shown in \fref{f.pol}. 
As is well known, the polarization is positive for $u$ quarks and negative for $d$ quarks. 
Without positivity constraints, a nonzero ratio can be extracted for $u$ up to $x \approx 0.8$ and for $d$ up to $x \approx 0.6$. 
With positivity constraints this is extended further up to $x \approx 0.85$ and $x \approx 0.7$ for $u$ and $d$, respectively.
Given the phenomenological interest in the behavior of $\Delta q/q$ as \mbox{$x \to 1$}~\cite{Brodsky:1994kg, Jimenez-Delgado:2013boa, Jimenez-Delgado:2014xza}, our simultaneous extraction of unpolarized and helicity PDFs including the $W$-lepton data provides the most reliable determination of the ratios to date.

The inclusion of the latest $W$ data also provides unambiguous signs for $\pub$ and $\pdb$, leading to a positive $\pub/\bar{u}$ and a negative $\pdb/\bar{d}$, matching their quark counterparts.
Without (with) positivity constraints, $\pub/\bar{u}$ can be distinguished from zero up to values of $x \approx 0.35$ ($x \approx 0.40$), while for $\pdb/\bar{d}$ it can be distinguished from zero up to $x \approx 0.35$ ($x \approx 0.45$).  
As with the asymmetry, the inclusion of positivity constraints makes little difference below $x = 0.1$ for both the quarks and antiquarks but reduces the uncertainties at larger $x$.

\begin{figure}[t]
\includegraphics[width=0.48\textwidth]{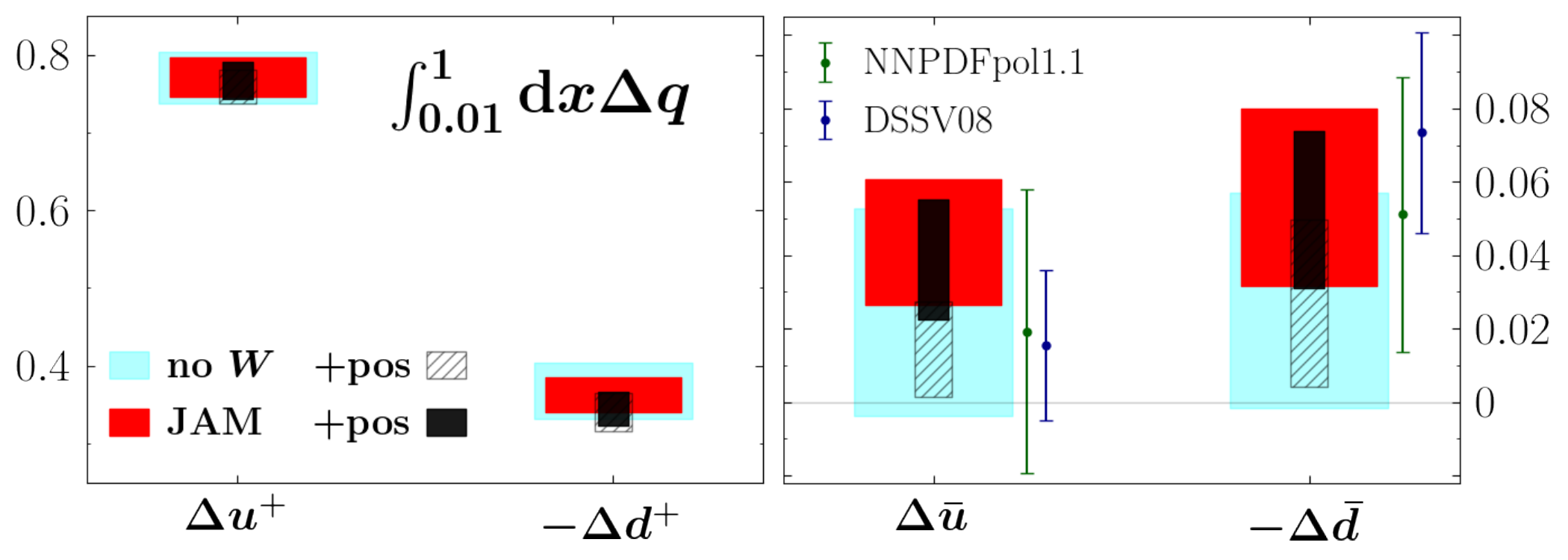}
\caption{Truncated integrals $\int_{0.01}^{1} \diff x\, \Delta q(x)$ at $Q^2=4$ GeV$^2$ for $\Delta u^+$, $-\Delta d^+$, $\Delta \bar{u}$ and $-\Delta \bar{d}$ from this analysis (red rectangles) compared with the fit without the RHIC $W/Z$ data (cyan) and with positivity constraints (small hatched squares without RHIC and black squares with RHIC). For the antiquarks, NNPDFpol1.1 (green points) and DSSV08 (blue points) are also shown.  The vertical height of the bands represents 1$\sigma$ uncertainty.}
\label{f.spin}
\end{figure}

Finally, in \fref{f.spin} we show the truncated integral $\int_{0.01}^{1} \diff x\, \Delta q(x)$ at $Q^2=4$ GeV$^2$ for the light quarks and antiquarks before and after including the RHIC $W$ data.
The lower limit of integration is chosen to roughly match the lower $x$ limit of the data. 
We see an improvement in the uncertainties for $\Delta u^+$ and $\Delta d^+$ of roughly 30\%, while $\pub$ sees an improvement of roughly 40\% and $\pdb$ an improvement of roughly 20\%.
While prior to the inclusion of the RHIC $W$ data the sign of the $\pub$ contribution to the proton spin was consistent with zero, after including these data we find that $\pub$ provides a small but unambiguously positive contribution to the proton spin.
Prior to the inclusion of the RHIC data, the result for $\pub$ depends heavily on the inclusion of positivity constraints.
When the RHIC data are included, however, this dependence is significantly reduced, allowing for an extraction that is far less dependent on theoretical assumptions.

Our truncated moments for $\Delta u^+$ and $\Delta d^+$, with values $0.771(25)$ and $-0.363(23)$, respectively, are only slightly smaller in magnitude than the corresponding full moments from lattice QCD calculations, which find $0.864(16)$ for $\Delta u^+$ and $-0.426(16)$ for $\Delta d^+$ \cite{Alexandrou:2020sml}. 
This comparison suggests that the contributions to the light quark moments below $x = 0.01$ are small.
We find nonzero truncated moments for $\Delta \bar{d}$ and, for the first time, $\Delta \bar{u}$, which was found to be consistent with zero in both the NNPDFpol1.1 and DSSV08 analyses. 
Interestingly, the contributions from $\pub$ [$+0.044(17)$] and $\pdb$ [$-0.056(24)$] approximately cancel in the sum.

{\it Outlook.---}\ 
Our analysis provides the first data-driven extraction of a nonzero polarized sea asymmetry 
at intermediate $x$
using the latest $W$-lepton data from RHIC, within a simultaneous global QCD analysis of polarized PDFs, unpolarized PDFs, and pion, kaon, and unidentified hadron FFs.
It also provides the first self-consistent extraction of the light quark polarizations and shows a nonzero contribution to the proton's spin from $\Delta \bar{u}$.

With the Jefferson Lab 12 GeV upgrade and the Electron-Ion Collider (EIC), future experiments will access new information on the spin structure of the proton \cite{Chang:2014jba, Geesaman:2018ixo}.
In particular, the high-luminosity CLAS12 SIDIS experiment using $K$ production~\cite{Hafidi:2008} will provide precise SIDIS data to complement the $W$-lepton production data from RHIC.
The EIC should bring forth new information on all polarized PDFs, in particular the strange and gluon PDFs~\cite{Zhou:2021llj}, while also extending the kinematic coverage of polarized DIS 
to lower $x$ and higher $Q^2$.


\begin{acknowledgments}
We thank F.~Ringer and W.~Vogelsang for the code used for calculating the $W$-lepton and jet cross sections, J.~J.~Ethier for his work on the $W$-lepton code and data collection, and P.~C.~Barry, P.~Schweitzer and Y.~Zhou for helpful discussions.
This work was supported by the U.S. Department of Energy Contract No.~DE-AC05-06OR23177, under which Jefferson Science Associates, LLC operates Jefferson Lab, and the National Science Foundation under grant number PHY-2110472.
The work of C.C. and A.M. was supported by the U.S. Department of Energy, Office of Science, Office of Nuclear Physics, within the framework of the TMD Topical Collaboration, and by Temple University (C.C.).
The work of N.S. was supported by the DOE, Office of Science, Office of Nuclear Physics in the Early Career Program.
This material is based upon work supported by the U.S. Department of Energy, Office of Science, Office of Workforce Development for Teachers and Scientists, Office of Science Graduate Student Research (SCGSR) program. 
The SCGSR program is administered by the Oak Ridge Institute for Science and Education (ORISE) for the DOE. ORISE is managed by ORAU under contract number DE‐SC0014664.
All opinions expressed in this paper are the author’s and do not necessarily reflect the policies and views of DOE, ORAU, or ORISE.
\end{acknowledgments}


\end{document}